\documentclass{aa}

\newcommand{\kms}{\ensuremath{\mathrm{km\,s^{-1}}}}
\newcommand{\ms}{\ensuremath{\mathrm{m\,s^{-1}}}}
\newcommand{\vsini}{\ensuremath{v\sin i}}

\newcommand{\teff}{\ensuremath{T_{\rm eff}}}
\newcommand{\logg}{\ensuremath{\log g}}                     

\usepackage{graphicx}
\usepackage{txfonts}
\usepackage{color}

\begin{document}

   \title{Stable magnetic fields and changing starspots on Vega}

   \subtitle{An ultra-deep decadal survey at Pic du Midi and OHP}

     \author{
        T. B\"ohm\inst{1,2,3}\and
        M. Holschneider\inst{4}\and
        P. Petit \inst{1,2,3}\and 
        F. Ligni\`eres\inst{1,2,3}\and
        F. Paletou\inst{1,2,3}\and
        C. P. Folsom\inst{5}\and
        M. Rainer\inst{\ref{inst:brera}} 
            }


   \institute{ 
        Universit\'e de Toulouse; UPS-OMP; IRAP; Toulouse, France\\
              \email{torsten.boehm@irap.omp.eu}\and       
        CNRS; IRAP; 14, avenue Edouard Belin, 31400 Toulouse, France\and
        Observatoire Midi-Pyr\'en\'ees, 14 ave Edouard Belin, 31400 Toulouse, France\and
        Institut f\"ur Mathematik, Universit\"at Potsdam, 14476 Potsdam, Germany\and
        Tartu Observatory, University of Tartu, Observatooriumi 1, T\~{o}ravere, 61602, Estonia
          \and
 	    INAF - Osservatorio Astronomico di Brera, via E. Bianchi 46, 23807 Merate, Italy\label{inst:brera}     
             }

   \date{Received July 8$^{th}$, 2025; accepted August 13$^{th}$, 2025}

 
  \abstract
   {}
   {Monitoring magnetic and activity variations in A and B stars with ultra-weak 
   magnetic fields is essential to understand the origin and evolution of these  
   fields in this domain of the HR diagram. Vega is a prototype star of it's category 
   and its long-term monitoring is most promising.}
   {High resolution spectrocopic and spectropolarimetric data were gathered with SOPHIE/OHP 
   in 2018 and NARVAL/NEO-NARVAL/TBL in 2018, 2023 and 2024. A total of 13108 individual spectra 
   of Vega were obtained and analyzed in this article. Magnetic field maps were reconstructed with 
   the Zeeman-Doppler imaging method and activity maps were reconstructed with an innovative code 
   built for this purpose. They were then compared to previous results.}
   {The rotation period of Vega was confirmed to be very close to the former reference period 
   of 0.678\,d. The average magnetic field confirms a negative spot of radial field on the pole with stable 
   strength, while the magnetic maps confirm the long-term stability of an oblique dipole, as well as smaller magnetic features showing up consistently throughout our three observing epochs. However, brightness maps show strong variations of the spot location over 
   a time-range of years, perhaps quicker. The spot contrast is very similar between $2012$, $2018$, $2023$ and $2024$, 
   with a normalized spectral amplitude of 0.0003. A direct correlation between 
   magnetic field and brightness 
   patches could not be revealed in the simultaneous SOPHIE and NARVAL $2018$ data set.}
   {Vega's "magnetic mystery" is getting more and more complex since indications point towards the presence of a fossil 
    magnetic field but also to the presence of a dynamo-generated field, most likely concentrated on equatorial regions.}


   \maketitle
%

\section{Introduction}

In 2009, the first important discovery of a weak magnetic field in a normal
A-type star (Vega, $-0.6 \pm 0.3$\,G for the disk-averaged line-of-sight
component) was announced by~\cite{vegamag}. Since then, this prototype star has
been studied extensively in spectroscopy, velocimetry, and spectropolarimetry.
Vega's characteristics and physical parameters are described in detail in~\cite{boehm2015}, 
the most important properties for this article being the recently redetermined quasi
pole-on vision of $i = 6.4\pm0.5^{\circ}$, a $\vsini = 21.6\pm0.3\ \kms$ 
and an equatorial velocity $v_{\rm eq} = 195\, \pm 15\, \kms$, based on an in-depth Fourier analysis
of Vega's spectral line profiles \citep{takedaparam}. 

Observations during five nights with SOPHIE/OHP in 2012 led~\cite{boehm2015} to
the major discovery and unambiguous confirmation of a structured stellar surface
at very faint amplitude levels in the dynamical spectra of $\Delta {\rm F/Fc}  \sim 5 \times 10^{-4}$ (Fc being the continuum flux)
corotating with the star. This first strong evidence that standard A-type stars
can show surface structures opens a new field of research and asks the question
about a potential link with the recently discovered weak magnetic field
discoveries in this category of stars.

In~\cite{petit2017}, based on the same data set, have been presented reconstructed Doppler
Imaging maps of the star. Maps corresponding to different nights are mostly
coherent, with apparent changes possibly due to the residual noise. Also,
marginal indications were found for a differential rotation of Vega, with the
pole and the equator rotating at higher rates than the intermediate latitude
regions. Still, the overall spectroscopic magnetic field signature (averaged
circularly polarized line profiles) and the derived surface magnetic maps for
the different epochs between 2008 and 2018 remained very similar, showing a
striking magnetic stability of Vega over one decade~\citep{petit2022}.

Following the first discovery of a magnetic field in Vega in 2009, similar fields 
were detected in other bright Am stars
(\citealt{petit2011},~\citealt{blazere2016}). The idea therefore gained support that very 
weak magnetic fields are present in many A and B type stars. 
A rotational modulation linked to the presence of star-spots might have been revealed 
in intermediate-mass stars lacking strong magnetic fields, the observations relying on 
Kepler photometry (\citealt{balona2017},~\citealt{balona2019}).

These results are of major importance for our knowledge of the evolution
of intermediate-mass stars. Before that, this category of stars with radiative
outer layers was not supposed to show magnetic field generation. In contrast to
lower mass stars with developed subphotospheric convection zones (like the sun),
no mechanism of magnetic field and associated stellar surface structure generation has
yet been established for this type of standard A-type star. Proposed scenarios include either
decaying weak fossil field~\citep{braithwaite2013} or dynamo fields constantly
regenerated either in the thin He II convective layer~\citep{cantiello2019} or
in the differentially rotating radiative envelope through Tayler
(\citealt{spruit2002},~\citealt{petitdemange2024}) or magneto-rotational
instabilities \citep{meduri2024}.

Studying Vega as a prototype star extensively via spectropolarimetric observations will provide a first understanding of the nature of the magnetic field, i.e. fossil and/or dynamo, or a combination of both. A varying magnetic field in Vega would suggest that a dynamo process is at work in the envelope of intermediate-mass stars.  At this stage we do not yet see an observational way of differentiating between the various amplification mechanisms possibly involved in a dynamo.

The direct comparison of successive magnetic field maps produced by the
Zeeman-Doppler imaging tomographic method, as well as spot maps produced in
unpolarized light by Doppler imaging or similar reconstruction techniques
enables us to constrain the origin and properties of the magnetic field, by
unveiling possible temporal changes related to dynamo action. Beyond the
question of intermediate-mass magnetism, the Vega Doppler maps also provide the
first detailed constraints on the flow structure at the surface of a stellar
radiative envelope.

Section \ref{obs} presents the observations and data reduction, Sect.~\ref{pol_res} 
presents the spectropolarimetric results, Sect.~\ref{act_res} presents the
analysis of the activity tracing starspots present in the different data sets.
Finally, a discussion and conclusion section follows in Sec.~\ref{disconc}.

\section{Observations and data reduction}\label{obs}

Table~\ref{table:log} summarizes the high-resolution observations analyzed in
this study. In 2018 Vega was observed simultaneously in spectropolarimetry with
NARVAL (TBL/Pic du Midi, France,~\citealt{auriere_narval}) and SOPHIE/OHP
(\citealt{perruchot2008},~\citealt{bouchy2013}). The goal was to combine stellar
magnetic field observations with intensity observations obtained with an
instrument stabilized in velocimetry - the intrinsic radial velocity variations
of SOPHIE being lower than 2-3 $\ms$.  While NARVAL acquired spectra in
polarimetric mode at $R= 65000$, SOPHIE was used in its high-resolution mode 
($R =75000$). The polarimetric results obtained with NARVAL in $2018$ are already part
of a first publication~\citep{petit2022}, while the intensity profiles were not
exploitable at that time due to data reduction issues of the former pipeline;
this was solved in this paper thanks to our new data reduction software
described below. In 2019 NARVAL was upgraded to the highly stabilized
spectropolarimeter/velocimeter NEO-NARVAL~\citep{boehm_neo}. The $2023$ and $2024$
data set observed with NEO-NARVAL suffered from poor weather conditions and a
slightly less efficient instrument, basically due to a loss of blue flux. As 
Table~\ref{table:log} shows a total of
13108 individual spectra of Vega were obtained and analyzed in this article. 
Additional archival data from our 2012 observing run was used; this data set is described in detail in \cite{boehm2015}.

For each run, the observing strategy was to obtain as many well exposed Vega
spectra during the whole night as possible in a continuous way, and to have all
allocated nights as closely spaced as possible (in the best case five
to seven nights following each other). Vega has a rotation period of 
$0.678^{+0.036}_{-0.029}\,$d detected in spectropolarimetric magnetic field
observations by~\cite{alina2012} and confirmed in surface intensity observations
by~\cite{boehm2015}. This short rotation period enables complete and dense rotational phase coverage with the presented observing strategy.

Data reduction of the SOPHIE data set was done with the SOPHIE Data Reduction
Software~\citep{bouchy2009}. The NARVAL and NEO-NARVAL data sets were reduced 
with the NEXTRA Data Reduction Software (Böhm, T. \& Holschneider, M., publication in prep.), 
for homogeneity. An alternative reduction of the NARVAL data with "Libre-Esprit", 
Donati et al. (1997) was also used for comparison, to ensure the quality of 
the NEXTRA data reduction. The NEXTRA
pipeline implements the flat-relative optimal extraction algorithm of \cite{Zechmeister_opt} 
and produces intensity Stokes I and Stokes V spectra,
based on the combination of four polarimetric sub-exposures.  All resulting
spectra were normalized to the local continuum by a dedicated normalization
routine within NEXTRA based on B-spline-fitting of each order at selected
wavelength positions, each star requiring an individual pointing file fixing
continuum zones. The spectra were also set in the barycentric velocity frame making use
of the python library barycorrpy~\citep{barycorrpy}. 

The next step was the extraction of least-square deconvolved (LSD) equivalent
Stokes I (unpolarised) and Stokes V (circularly polarised) photospheric line
profiles (\citealt{donati1997},~\citealt{kochukhov2010}) for all spectra. The
underlying idea of the LSD technique is to make use of the multiplex information
of hundreds to thousands of spectral lines in order to produce a common very
high $S/N$ equivalent photospheric line profile. This applies also for LSD
Stokes V profiles taking into account the effective Land\'e factor of each line
revealing the sensitivity of the line to magnetic fields.  An atomic data
linelist was used with  $\teff = 10000$\,K and $\logg = 4.0$, yielding a spectral line mask corresponding to Vega's polar temperature and surface gravity using solar abundances. Eventually, only $207$ spectral lines were retained using the
interactive tool SpecpolFlow (\citealt{specpolflow1},~\citealt{specpolflow2}), based
on the best fitting of mostly unblended spectral lines in normalized continuum
zones excluding hydrogen lines, "flat-bottomed" lines (see~\citealt{takeda2008}),
and telluric line areas. This preparation was crucial to obtain high S/N LSD
profiles for Stokes I and Stokes V using LSDpy\footnote{Folsom,
https://github.com/folsomcp/LSDpy}. Velocity
sampling was adapted to the slightly different resolution of SOPHIE and NEO-NARVAL
(SOPHIE: $1.7\,\kms$ versus NEO-NARVAL: $2.3\,\kms$). Theoretically implementing a 
super-resolved oblique PSF extraction could mitigate the larger pixel size (see e.g. \cite{2021A&A...646A..32P}), but doing so would impact the noise properties 
and makes estimating reliable uncertainties challenging, so we chose not do 
to this in NEXTRA. Normalisation of the LSD profile
calculation was done using a profile depth of 0.31, an effective land\'e factor of 1.22 and a wavelength of $470\,nm$. While intensity LSD profiles were weighted with the line depth, Stokes V LSD profiles used depth $\times$ wavelength $\times$ effective land\'e factor (g) for weighting the spectral lines.

\begin{figure} \centering
\includegraphics[width=9cm]{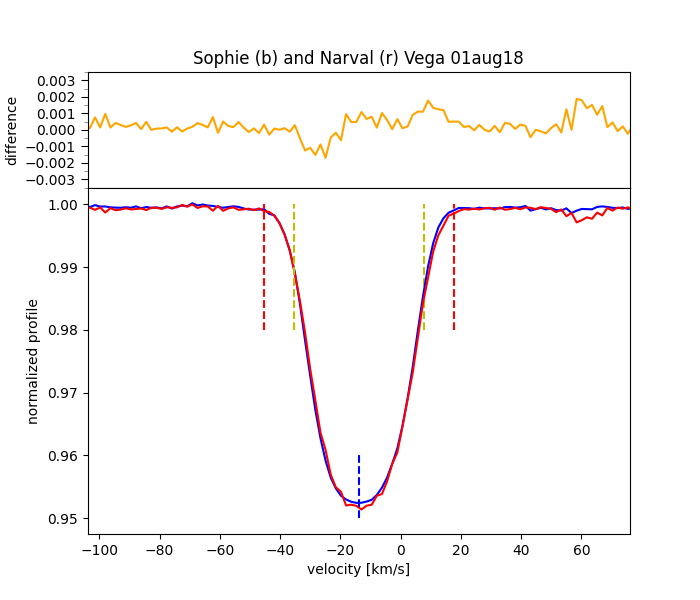} \caption{\small
Comparison of the LSD Stokes I intensity profile of Vega obtained with NARVAL/TBL  and the new NEXTRA pipeline (red) and a quasi simultaneous observation with SOPHIE/OHP (within one minute, blue), both processed with LSDpy. Stokes I profiles are quasi identical. The difference of the SOPHIE and NARVAL  profile is  presented in the top panel (orange). Dashed blue line corresponds to the radial velocity of Vega of
\mbox{$-13.9\,\kms$}. The dashed yellow lines show $\pm\vsini$ \,($\mbox{21.6}\,\kms$), while the dashed red lines correspond to the outer limits of
the profile including gaussian broadening (at additional $\pm \mbox{10}\,\kms$).
Data are from August 1$^{\rm st}$ 2018 at 23:11 UT.} \label{NEXTRA_I}		
\end{figure}

\begin{figure} \centering
\includegraphics[width=9cm]{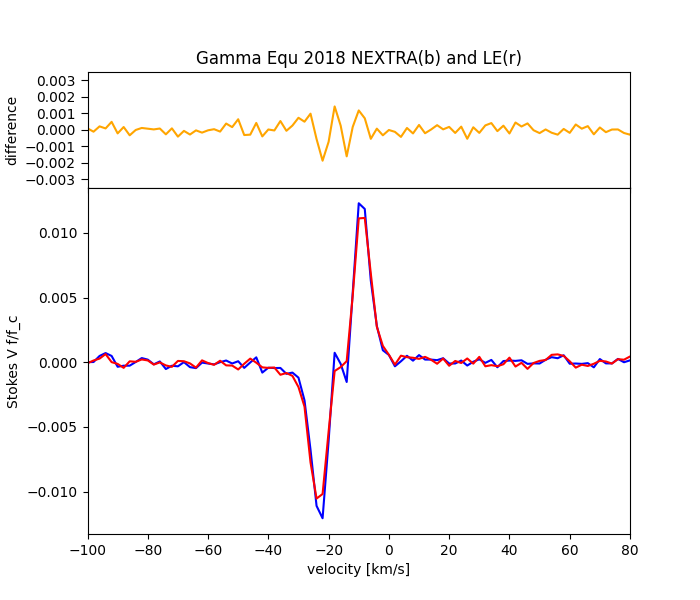} \caption{\small
Comparison of the LSD Stokes V profile of $\gamma$
Equ generated with LSDpy and reduced with the new NEXTRA spectropolarimetric
data reduction pipeline (blue) and reduced with the historic
"Libre-Esprit"~(\cite{donati1997}, red), Both pipelines provide quasi identical
results. The exact same data set from August 19$^{\rm th}$ 2018 is compared. The difference of the NEXTRA and LE profile is presented in the top panel (orange).}
\label{NEXTRA_V}		
\end{figure}

The accuracy of this new NEXTRA pipeline is shown in Fig.~\ref{NEXTRA_I}
and Fig.~\ref{NEXTRA_V} and will be detailed in a forthcoming article (B\"ohm,
T. \& Holschneider, M., publication in prep.). The SOPHIE, NEXTRA and LE
("Libre-Esprit",~\citealt{donati1997}) pipelines yield, when applicable, highly
comparable Stokes V and Stokes I profiles on the different data sets.

\begin{table*} \caption[]{Log of the spectroscopic and spectropolarimetric
observations of Vega.}
\label{table:log} 
\centering 
\begin{tabular}{cccccccc} \hline\hline Year   &
Instrument   & BJD$_{\rm first}$	& BJD$_{\rm last}$	& t$_{\rm cov}$ (d)  &
N$_{\rm spec}$	&  t$_{\rm exp}$ (s) & S/N \\ (1)		&  (2)  	  &  (3)        		 &
(4)                      &        (5)        		&     (6)             &   (7)
& (8)   \\ \hline 2018    &   Narval         &   8331.3406        &   8337.6238
&      6.28            	& 568          &        4x13      & 2216$\pm$326     \\
2018    &   SOPHIE      &   8331.3418        &   8336.6512         &     5.31
&2704          &         20        & 934$\pm$324     \\ 2023    &   NEO-NARVAL
&  10152.3752       &  10169.5444       &    17.12    		& 1426          &
4x15      & 1242$\pm$536     \\ 2024    &   NEO-NARVAL  &  10564.3111        &
10571.4165       &     7.11 			&  607          &         4x15       &
747$\pm$179     \\ \hline \end{tabular}
\tablefoot{ (1) Year of observations; (2) Instrument (Narval(TBL):
spectropolarimetry, NEO-NARVAL(TBL): spectropolarimetry, SOPHIE(OHP)
spectroscopy); (3) and (4) Barycentric Julian date (mid observation, 2450000+)
of the first and the last stellar spectrum of the run; (5)  BJD range; (6)Number
of high resolution intensity spectra obtained - for NARVAL and NEO-NARVAL the
total number of acquired spectra is 4 times the amount indicated in this column,
an intensity spectrum being the combination of 4 subexposures; (7) exposure
time for SOPHIE or individual exposure times in a polarimetric sequence for NARVAL/NEO-NARVAL;  (8) Mean and standard deviation of the signal to noise ratio per resolved element at 520nm.}

 \end{table*}

\section{Spectropolarimetric results}\label{pol_res}

Once the data reduction was performed with the respective pipeline for SOPHIE and
NARVAL/NEO-NARVAL, all spectra were processed in the exact same way in order to
obtain the LSD profiles described in Section \ref{obs}. Based on the noise
level in the Stokes Null profiles a certain quantile of files was kept,
rejecting the most noisy onces. At the end most analysis were done with a
rejection level above the 0.95 quantile. 

\subsection{Average Stokes V profiles}

\begin{figure} \centering
\includegraphics[width=10cm]{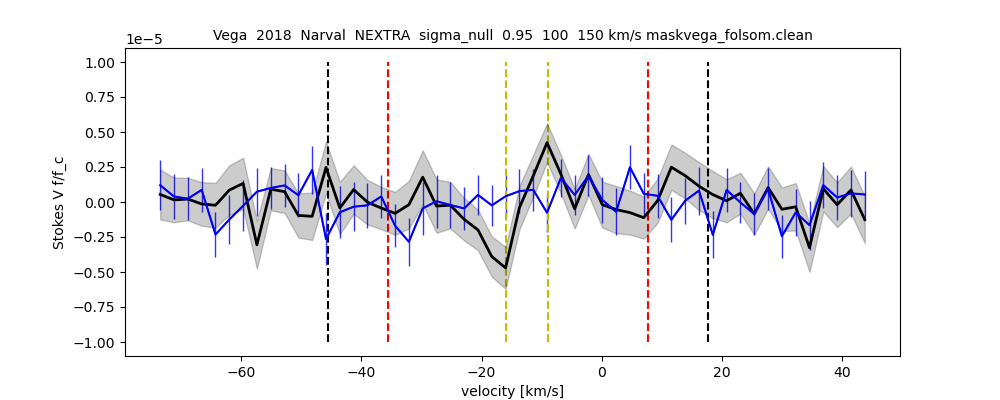}
\includegraphics[width=10cm]{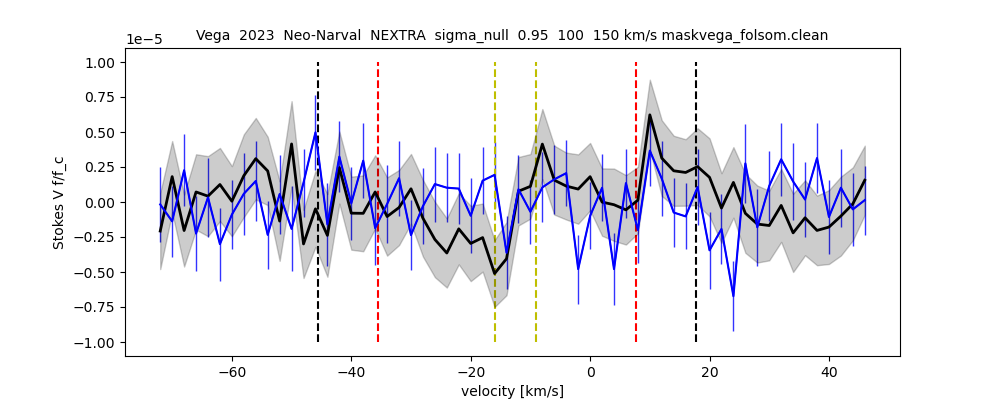}
\includegraphics[width=10cm]{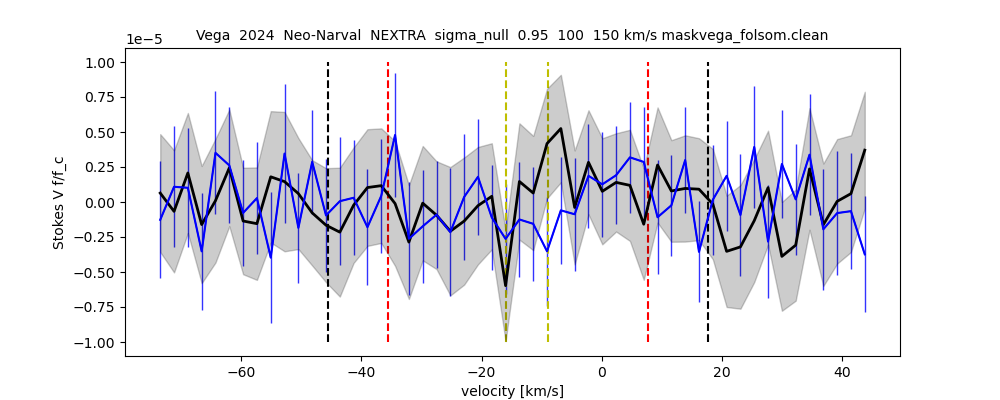}
\caption{\small  Average Stokes V profile of Vega: 2018 (top), 2023 (middle),
2024 (bottom).  Black line and grey zone: Averaged Stokes V and $\pm
1\,\sigma$ envelope. Blue line and $1\,\sigma$ error bars: Averaged Null
profile. Vertical dashed yellow lines show the position of the negative and
positive peak of the 2018 averaged Stokes V profile. The dashed red lines show
$\pm\vsini$, while the dashed black lines correspond to the outer limits of the
profile including gaussian broadening (additional $\pm\,10\,\kms$) as shown in
Fig.~\ref{NEXTRA_I}.}\label{vega_average}		
\end{figure}

\begin{figure} \centering
\includegraphics[width=10cm]{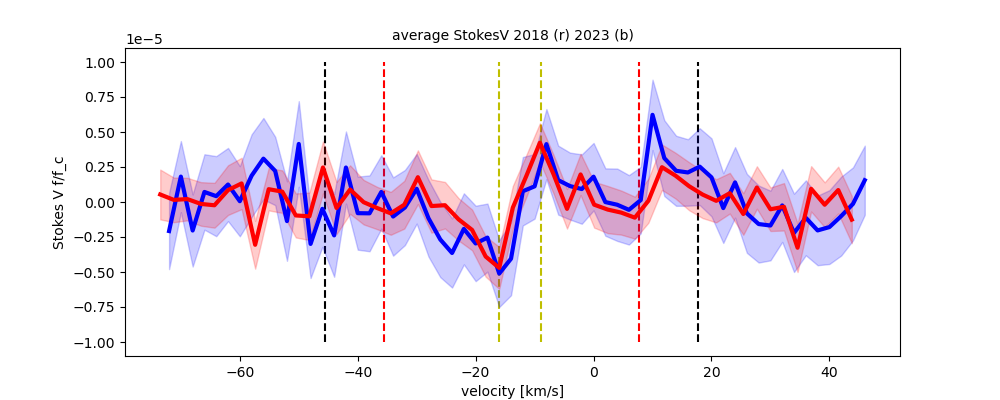}
\includegraphics[width=10cm]{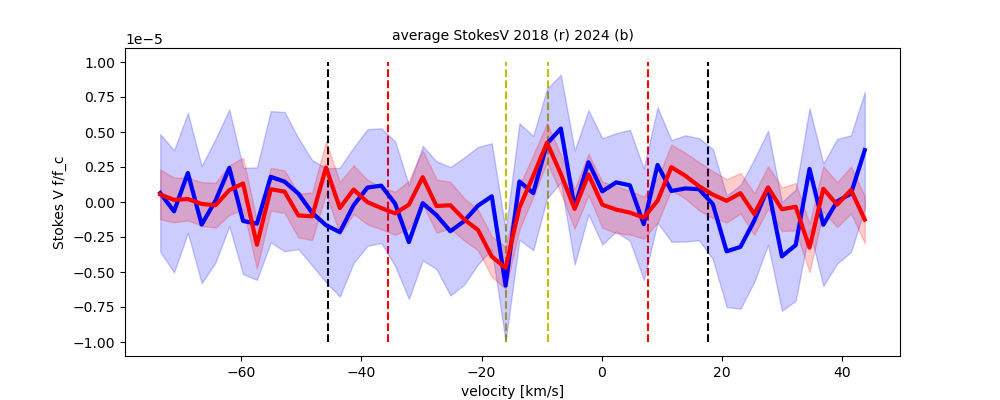}
\caption{\small Comparison of the average LSD Stokes V profile from NARVAL in 2018 (red) with profiles from NEO-NARVAL  in 2023 (top panel, blue) and in 2024 (bottom panel, blue). The vertical lines are
described in  Fig.~\ref{vega_average}.}\label{vega_comp}
\end{figure}

A first step of the analysis was to produce the averaged Stokes V profiles by combining all profiles of each run. 
Fig.~\ref{vega_average} shows three averaged Stokes V profiles, which appear to be consistent, indicating that this measure does not seem to change over a decade. 
As a reference value are plotted the averaged null profiles. Fig.~\ref{vega_comp} shows a direct comparison of the different profiles with the 2018 data set. 
What can be seen is the narrow velocity width of the  averaged Stokes V signal. 

This implies that the the average stable Stokes V profile is produced by a component of the magnetic field that is concentrated very close to the rotation pole. Since these average profiles show the same shape as in~\cite{petit2022}, polar magnetic 
field strength seem to remain constant over the years.

\subsection{surface magnetic geometry}

\begin{figure*}
\centering
\includegraphics[width=18cm]{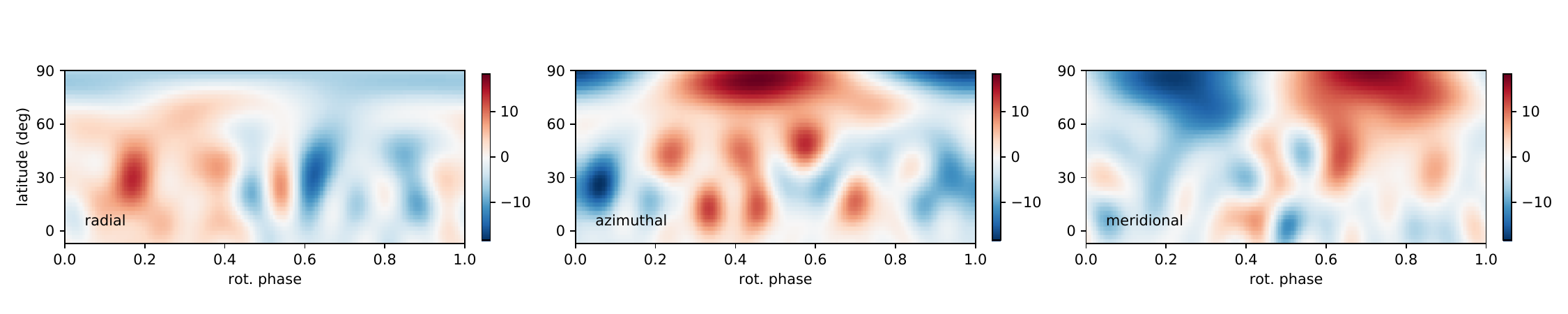}
\includegraphics[width=18cm]{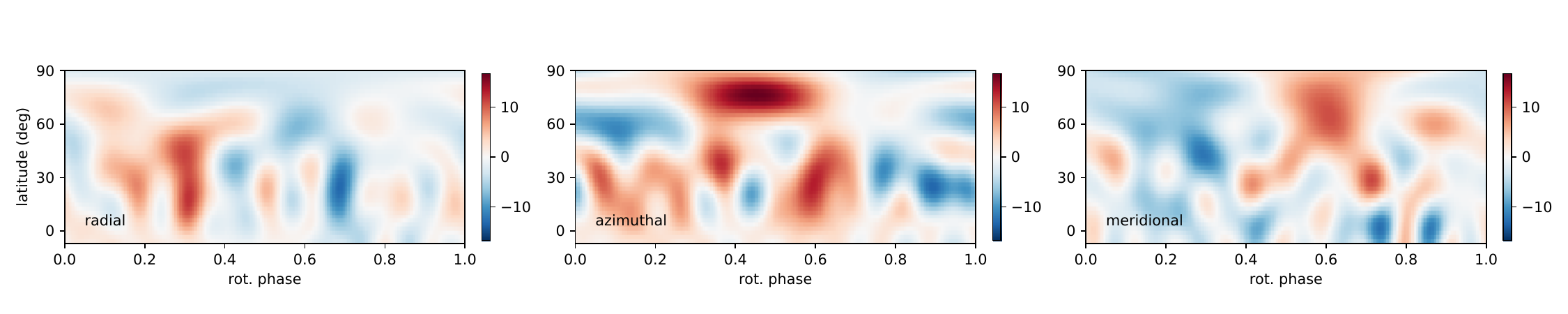}
\includegraphics[width=18cm]{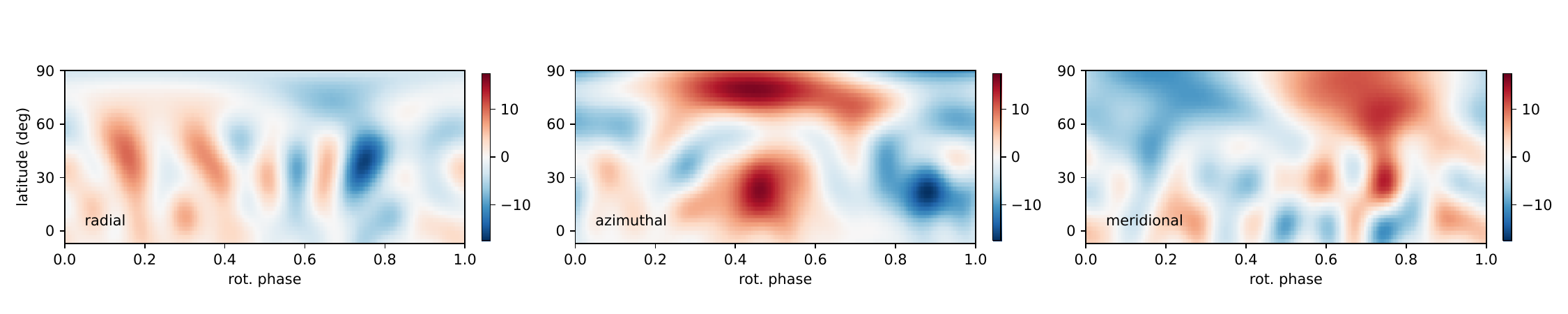}
\caption{From top to bottom, ZDI reconstructions of magnetic geometries in equatorial projection 
for 2018, 2023, and 2024. Each column displays a component of the local magnetic vector, in a spherical 
coordinates frame. The field strength is color-coded (Gauss unit). The phase reference is taken at BJD = 2456892.015, 2456892.185, and 2456892.506 for 2018, 2023, and 2024 (respectively), to ensure that all epochs display an oblique dipole pointing at the same phases.}
\label{magneticmaps}		
\end{figure*}

The successive time-series of Stokes V LSD profiles were used to reconstruct 
the surface magnetic field distribution of Vega at different epochs. For this, 
we employed the Zeeman-Doppler Imaging (ZDI) inversion method~\citep{1989A&A...225..456S}, 
using the Python code of~\cite{2018MNRAS.474.4956F}, based on  the spherical harmonics 
formalism of~\cite{2006MNRAS.370..629D}, the local line model of~\cite{2018MNRAS.481.5286F}, 
and assuming an oblate surface shape \citep{2020A&A...643A..39C}. 
The data preparation and ZDI input parameters were chosen identical to those adopted by~\cite{petit2022}
 for earlier observations of Vega, except that we adopted a rotation period of 0.678~d. The resulting reduced $\chi^2$ values are equal to 0.67, 0.55, and 0.28 in 2018, 2023, and 2024 (respectively). Three magnetic maps for observations taken in 2018, 2023, and 2024 are shown in Fig.~\ref{magneticmaps}.

Data gathered in 2018 were already presented by~\cite{petit2022}, although  the reduction 
code and LSD pipeline were different. Given the extremely weak signatures involved, it is 
interesting to note that the main magnetic features highlighted by \cite{petit2022} survived 
this independent data reduction. The first one is the patch of radial negative field close to 
the pole, and the second one is the equator-on dipole (the dominant reconstructed feature, storing alone 30 to 50 \% of the poloidal magnetic energy, depending on the epoch). Other small-scale magnetic patches show less consistency with our previous 
reconstruction, although some medium-sized magnetic spots do show up in both maps. The two newer maps, obtained from previously unpublished data of 2023 and 2024, 
confirm the long-term stability of the polar spot and inclined dipole. These characteristics are 
consistently recovered from data sets collected since 2008 (but do not show up if the Null profile 
is used instead of the Stokes V profile), enabling us to estimate a lifetime of 16 years, at least.

We have to deal here with observations where Zeeman signatures in individual LSD Stokes V profiles are dominated by noise, so that hundreds to thousands of LSD profiles must be combined in the tomographic inversion to extract a rotationally-modulated signal. In such extreme situation, it is difficult to ensure that the tomographic inversion 
does not suffer from overfitting. In other words, it is difficult to make sure that part of the recovered 
magnetic distribution does not originate from an attempt of the ZDI code to fit the noise pattern, 
especially at small spatial scales (as already stressed in \citealt{petit2022}). This limitation is critical here, since rotationally-modulated spectral features seen in Stokes~I LSD 
profiles seem to be linked to surface regions smaller than the oblique dipole. The fact that small 
magnetic spots are not fully consistent in 2018, depending on the reduction code (i.e. between the map shown here and the one of \citealt{petit2022}), is a further
confirmation that small features should be considered with care, except for the polar spot 
which benefits from an optimal visibility throughout the rotation cycle. Splitting 2018 observations 
in two subsets to reconstruct two maps (either by taking every second observations, or separating 
the first half from the second half of the observing epoch) lead to similar conclusions. In spite of this warning, our magnetic maps suggest that some of the smaller magnetic regions do exhibit some consistency between epochs. The poles of the oblique dipole seem to co-exist with a higher geometrical complexity. This is the case, e.g., of the two spots of positive radial field seen in 2024 at phases 0.2 and 0.4 (inside the positive pole of the dipole), while similar features were visible on previous years as well.

\section{Activity tracing starspots results}
\label{act_res}

Section \ref{rotation_period} presents a redetermination of the rotation period based on rotational modulation of spectroscopic lines, section \ref{dynamic} analyses starspot signatures in dynamic spectra of all data sets correpsponding to the different observing seasons, section \ref{spot_map} presents a newly developed reconstruction procedure for brightness maps and applies it to Vega, and finally \ref{search_corr} summarizes the search for correlation between magnetic field and activity structures.

\subsection{Rotation period redetermination}
\label{rotation_period}

As mentioned earlier, a first determination of the rotation period of Vega yielded $0.678^{+0.036}_{-0.029}$\,d. 
It was detected in spectropolarimetric magnetic field observations \citep{alina2012} and confirmed by surface 
intensity observations~\cite{boehm2015}. 

In a next step, as described in \cite{boehm2015}, the bisector of the Stokes I LSD profiles was determined. The bisector velocity span ($v_{\rm span}$) was calculated for each profile, measuring the difference of the upper and lower part of the bisector (a kind of skewness). For this, we worked in relative profile height, the bottom of the profile set at 0., the continuum at 1. $v_{\rm span}$ was calculated as the difference of the medians of upper [0.35, 0.5] and lower [0.1, 0.25] bisector ranges.

\begin{figure}
\centering
\includegraphics[width=10cm]{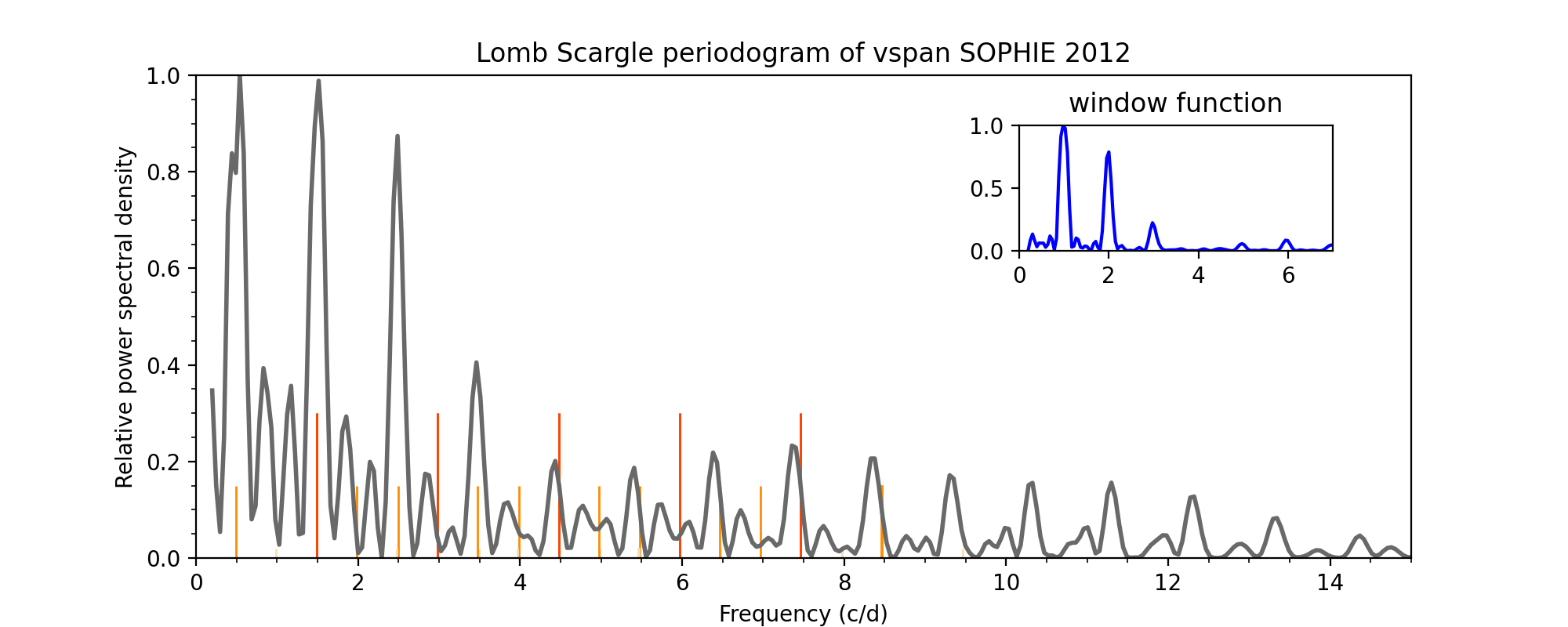}
\includegraphics[width=10cm]{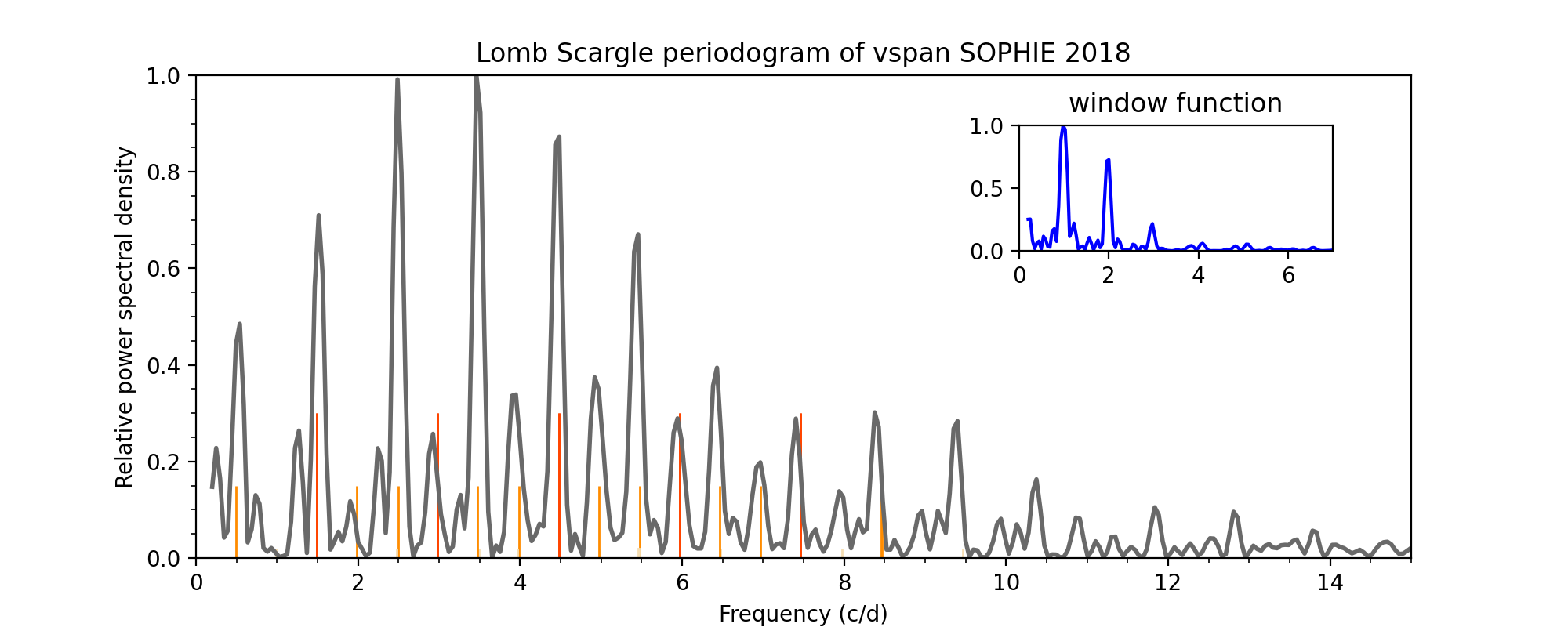}
\caption{\small Lomb Scargle periodogram of $v_{\rm span}$ (SOPHIE 2012 (upper panel), 2018
(lower panel)), including the window function of the data set. The rotational frequency of the star (1.492\,d$^{-1}$,  for a period of 0.670\,d), as well as its harmonics, are indicated by vertical bars with a height of 0.3. The lower bars at 0.15 and 0.02 height indicate the position of the $\pm1$ and $2-$day
aliases, respectively, that are generated by the window function.}
\label{ls_vspan_2018}		 
\end{figure}

Under the assumption of a constant rotation period over the
 timescale of 6 years, and based on the same measure ($v_{\rm span}$), the new 2018 SOPHIE data set enabled us to improve the 
 precision of the period by combining with the results from SOPHIE 2012 data set: a Lomb-Scargle analysis of the $v_{\rm span}$ 
 yields a rotation period of 0.6771$\pm 0.0023$ d, while a FELIX (\citealt{charpinet_felix}, 
 \citealt{zong_felix1}, \citealt{zong_felix2}) $v_{\rm span}$ analysis based on multiperiodic non linear 
 least square fit produces an averaged rotation period of $0.6705 \pm 0.0019$\,d (2012: $0.6679 \pm 0.0016$\,d; 2018: $0.6730\pm 0.0001$\,d).
  Both determinations are within a 3 sigma interval from the original $0.678^{+0.036}_{-0.029}$\,d period, 
  but an amelioration of the error bar lets us now assume a period of $0.6705\pm 0.0019$\,d as produced by 
  the FELIX time series analysis code. Fig.~\ref{ls_vspan_2018} shows the frequency analysis of the $v_{\rm span}$ 
  for the 2012 and 2018 data set, superimposed are the redetermined rotation period.
It should be noted that slight differences in rotation period determination may
be a consequence of Vega likely not being in solid body rotation, and different tracers might sense different areas of the star.

\subsection{Starspot signatures in dynamic spectra}\label{dynamic}

After refining the rotational period of Vega we produced 2D dynamical spectra, as described in \cite{boehm2015}.

\begin{figure*} \centering
\includegraphics[width=8cm]{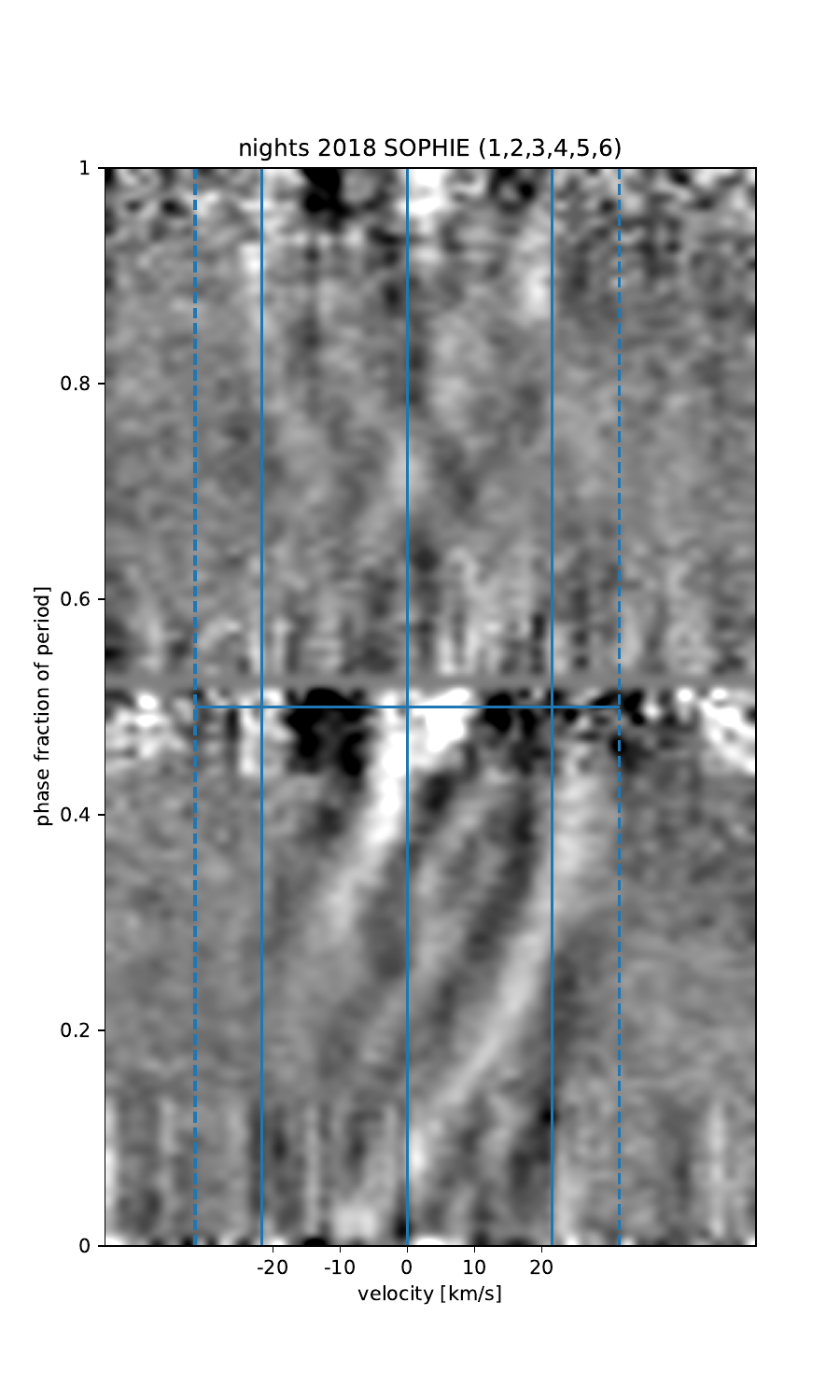}
\includegraphics[width=8cm]{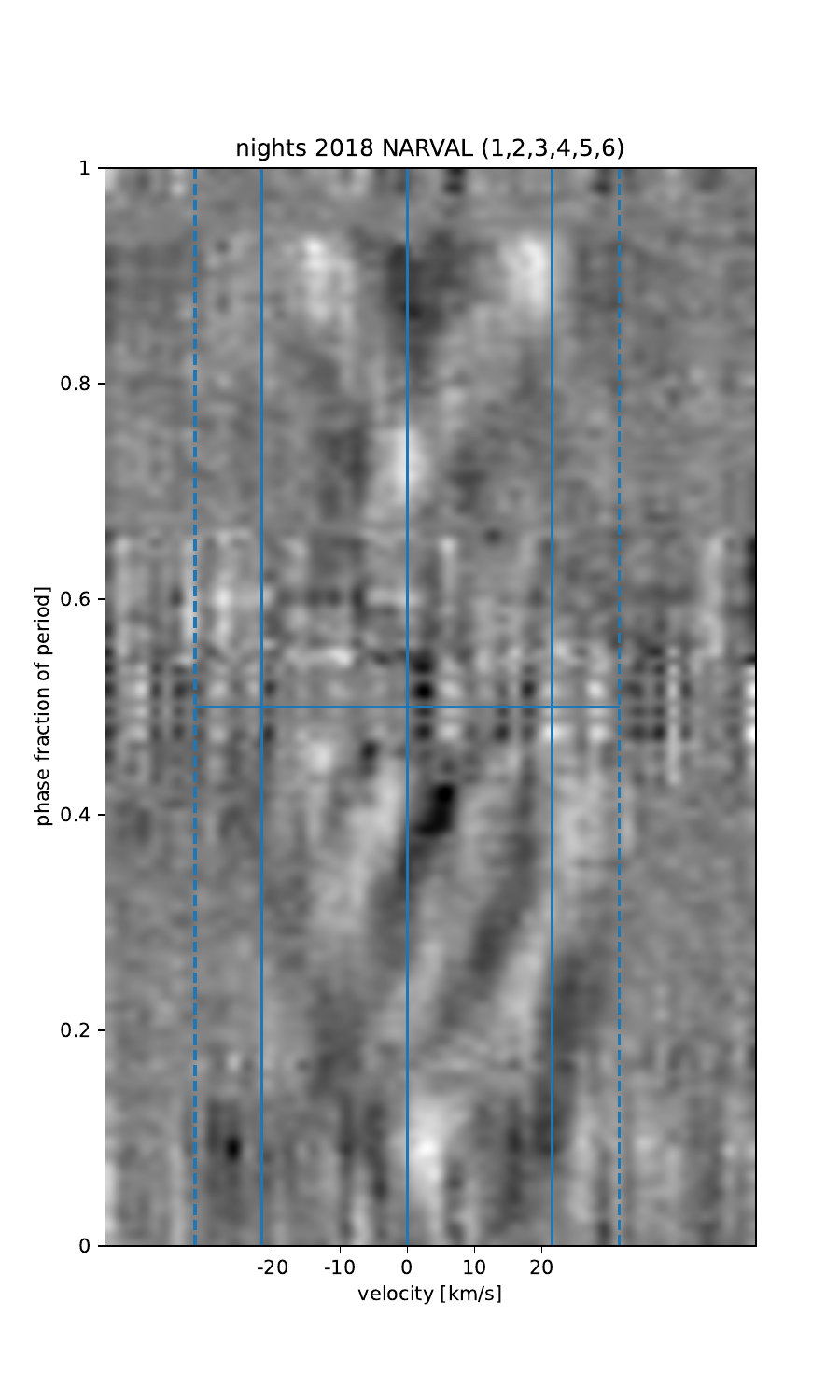}
\caption{\small Comparison of simultaneous 2018 Vega observations obtained with 
SOPHIE/OHP (left) and NARVAL/TBL (right). The two inner blue lines symmetric about zero velocity correspond to $\pm\,\vsini= 21.6\,\kms$
(the profiles have been recentered to the rest velocity),
while the outer dashed blue lines include the zone where gaussian broadening widens the line profile. 
Only $80\%$ of the highest $S/N$ files were kept for the dynamical spectra.  White signatures designate 
bright spots on the star, grey scale varies from -0.0002 (white) to +0.0002 (black), values outside 
this scale are saturated in the figure. Phase zero corresponds to $BJD = 2456892.015$.}\label{vega_2018_dynamic_comp}		
\end{figure*}

To do so, we made use of the LSD I profiles
described in Sect. \ref{obs}. All individual observing times were rephased with
a rotational period of P= 0.678\,d (for a sake of coherence with the magnetic
maps) with respect to the same time reference used in the magnetic maps (For 2012 data phase zero corresponds to BJD$_{\rm ref}$= 2456142.332, for 2018, 2023 and 2024 BJD$_{\rm
ref}$ = 2456892.015, 2456892.185, and 2456892.506, respectively). Therefore, for a given observing run, all magnetic and activity maps in this article can directly be compared. All LSD profiles were attributed to 128 phase bins and an
individual median profile was calculated for each phase bin.  In order to
exhibit the very faint structures in the dynamic profile representation of Fig.
\ref{vega_2018_dynamic_comp}, a mean profile was calculated throughout all
spectra of a night, and variations in line depth, radial velocity shifts of the
profile, as well as offset variations (normalization errors) were modeled via a
linear model. The difference between the observations and an individual three
parameter adjustment is then presented in the figure, as a function of velocity
recentered on the stars rest frame (Vega having a radial velocity of
$-13.9\,\kms$). We do not know at this stage the origin of the starspots observed in 
Vega~\citep{boehm2015}. These might be differences in brightness, but could also be related 
to local abundance variations, or a combination of both. If brightness related, a bright 
spot on the rotating star produces a local negative
dip in the spectroscopic profile. According to the grey scale, a bright signature in the dynamical spectrum corresponds to a bright stellar spot. The typical brightness
scale of dark and clear structures are $\pm 1.5 \times 10^{-4}\, {\rm F/F_{c}}$.

As a first step we wanted to insure that two different data sets obtained
simultaneously, i.e.~covering the same nights, and reduced with different data
reduction pipelines reproduce these very faint signatures in the same way. 
Fig.~\ref{vega_2018_dynamic_comp} shows two dynamical Vega spectra from 2018 obtained
with high precision velocimeter SOPHIE/OHP (left) and the spectropolarimeter
NARVAL/TBL (right). Both figures cover the same observing period and have a
starting period differing only by 1.6 minutes. SOPHIE data were reduced with the
dedicated Sophie pipeline and NARVAL data with the new NEXTRA pipeline, while
LSD profiles were produced in the identical way with LSDpy (see Sect.~\ref{obs}). 
We can see that the observations done with SOPHIE/OHP and NARVAL/TBL
present exactly the same features, despite the fact that SOPHIE has a slightly
higher spectral resolution, and confirm therefore the reality of these very
faint signatures.

The main question of this section concerns the variability of surface brightness spots 
on Vega (assuming brightness/temperature is involved), while surface magnetic field variations 
are analyzed in Sect.~\ref{pol_res}. Dynamical spectra obtained 
with SOPHIE/OHP in 2012 (left) and 2018 (right) are compared in 
Fig.~\ref{vega_2012_2018}. Since the precision on the 
rotation period is too low to directly compare rephased dynamical spectra obtained with a time 
difference of roughly 6 years, a comparison must be made based on the shape of different features. 
To better understand these signatures it should be understood that a vertical signature close to 
the zero velocity line corresponds to an almost polar spot, while a sinusoidal signature crossing from the negative to positive velocity extreme (approx. $\pm 30\,\kms$) indicates
a surface feature very close to the stellar equator. While 2012 bright surface spots seem to cover 
all velocity domains, 2018 bright spots seem to cover dominantly equatorial zones of the stellar surface. 
This can be seen for instance by comparing the average inclination of signatures crossing the profile 
or their velocity amplitudes. The signatures of the 2018 dynamic spectrum are systematically more 
inclined indicating a localization nearer to the equator. Also, certain features present in the 2018 plot, 
e.g.~the bright signature around phase 0.4 and maximum velocity are not seen at any phase in 
the 2012 dynamical spectra. 

The contrast of the spectral signature (difference between brightest and darkest part of the signature) 
is very constant over the years 2012, 2018, 2023 and 2024, with a total maximal amplitude of 0.0003 with 
respect to the normalized continuum. 

The value of 0.0005 in the~\cite{boehm2015} has been refined in this paper to 0.0003 thanks to improved data reduction and rejection of poor quality profiles.

\subsection{Reconstruction of Vega's brightness map}\label{spot_map}

\begin{figure*}
\includegraphics[width=16cm]{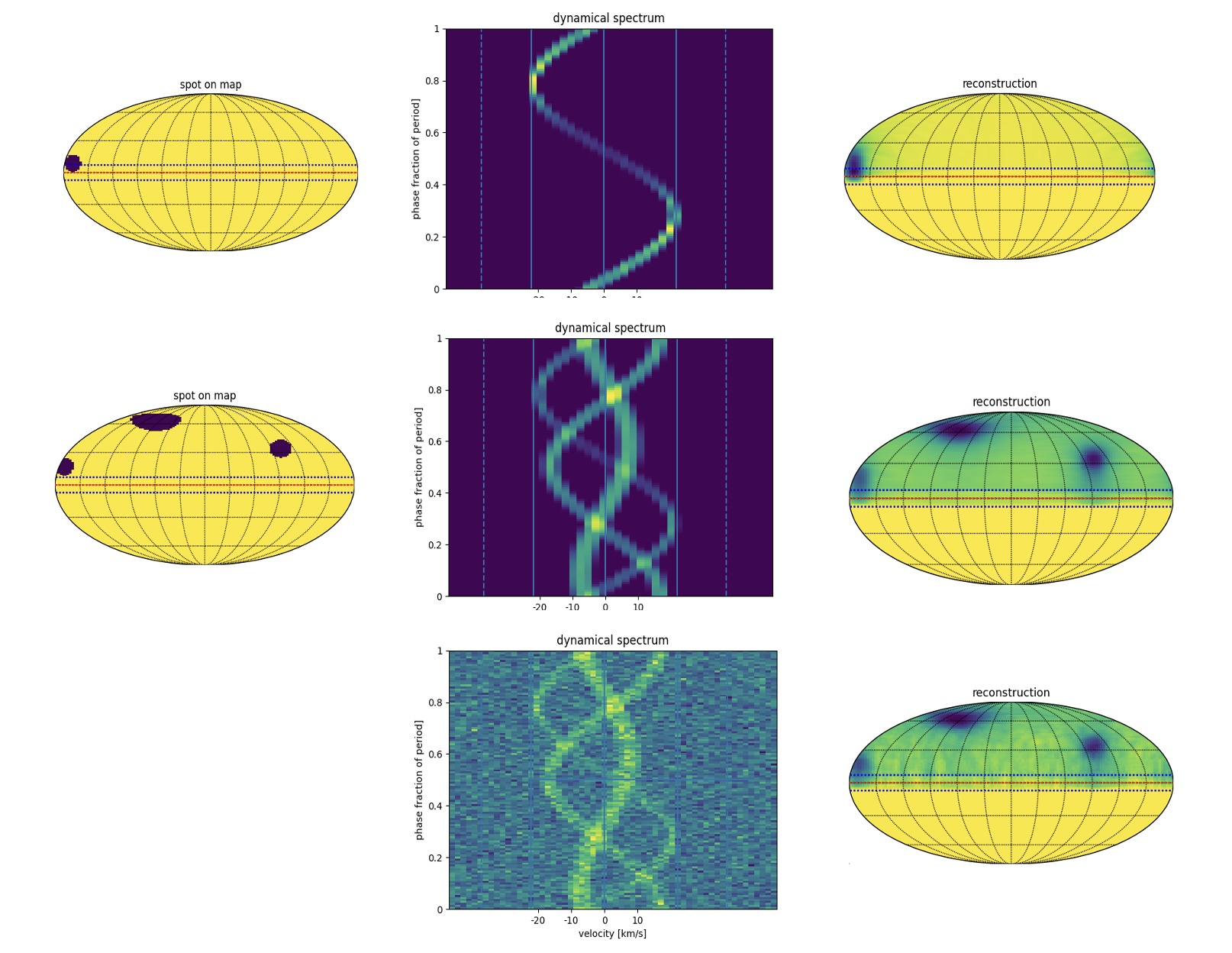}
\caption{\small  Simulation of a spotted stellar surface in Mollweide projection 
and seen almost pole on as Vega. Deep blue are dark spots which appear in emission in the dynamical spectrum. 
Phase $0$ is at the left of the stellar map (rotation vector points towards the top). 
Stellar rotation is counterclockwise in this representation. A spot at the central meridian of 
the stellar sphere (in front of the observer) would have phase zero and appear on the left of the map. 
Phases increase clockwise on the stellar sphere. If we read the map in analogy to the terrestrial coordinates, then the phase 
would correspond to the negative longitude. 
Left figure is the original map of the stellar surface, middle figure the dynamical spectrum, right figure shows 
the reconstructed stellar surface map. Upper: a single spot, middle: more complex pattern of three spots of different sizes. 
Lower: Simulation of a realistic dynamical spectrum (signatures with an amplitude of $1.5 \times 10^{-4}\, {\rm F/F_{c}}$ and a noise of $\sigma = 5\times 10^5{-5}$ .
}
\label{app_simul}	
\end{figure*}

We use a pixel based pattern matching approach to asses the spot density from the observed dynamical spectra. As can be seen in Fig. \ref{app_simul} each spot on a stellar surface shows a characteristic signature in the dynamical spectrum. Our basic reconstruction relays on this fact. To do so, We use a regular $128\times128$, colatitude $\times$ longitude grid over the (simplified spherical) star. Each cell of this grid will define a trace in the radial velocity $\times$ rotation phase plane, i.e. the dynamical spectrum. We then sum the amplitude of the observed 
intensity along the trace specific to each grid-point we want to reconstruct. This provides us with a measure for the signed spot-intensity at this at this particular location of the stellar surface. This simple reconstruction of the spot distribution includes an amplitude bias related in first order to its colatitude. We therefore divide the reconstruction by this colatitude dependency to reduce this bias. However, the "spread" bias due to the crossing of different signatures in the dynamical spectrum is not reduced that way. An application of the method is shown in the simulation of Fig. \ref{app_simul},

including a realistic simulation with signature amplitudes and gaussian noise corresponding to our observations. 
Phase 0 is at the left of the stellar map (the rotation vector points towards the top), phase 1 on the right. Stellar rotation is counterclockwise in this representation. A spot at the central meridian of the stellar sphere (in front of the observer) would have phase zero and appear on the left of the map. Phases 
increase clockwise on the stellar sphere. 
In our figures we adopted exactly the same color upper and lower boundaries for the color scale.

\begin{figure*}
\centering
\includegraphics[width=8cm]{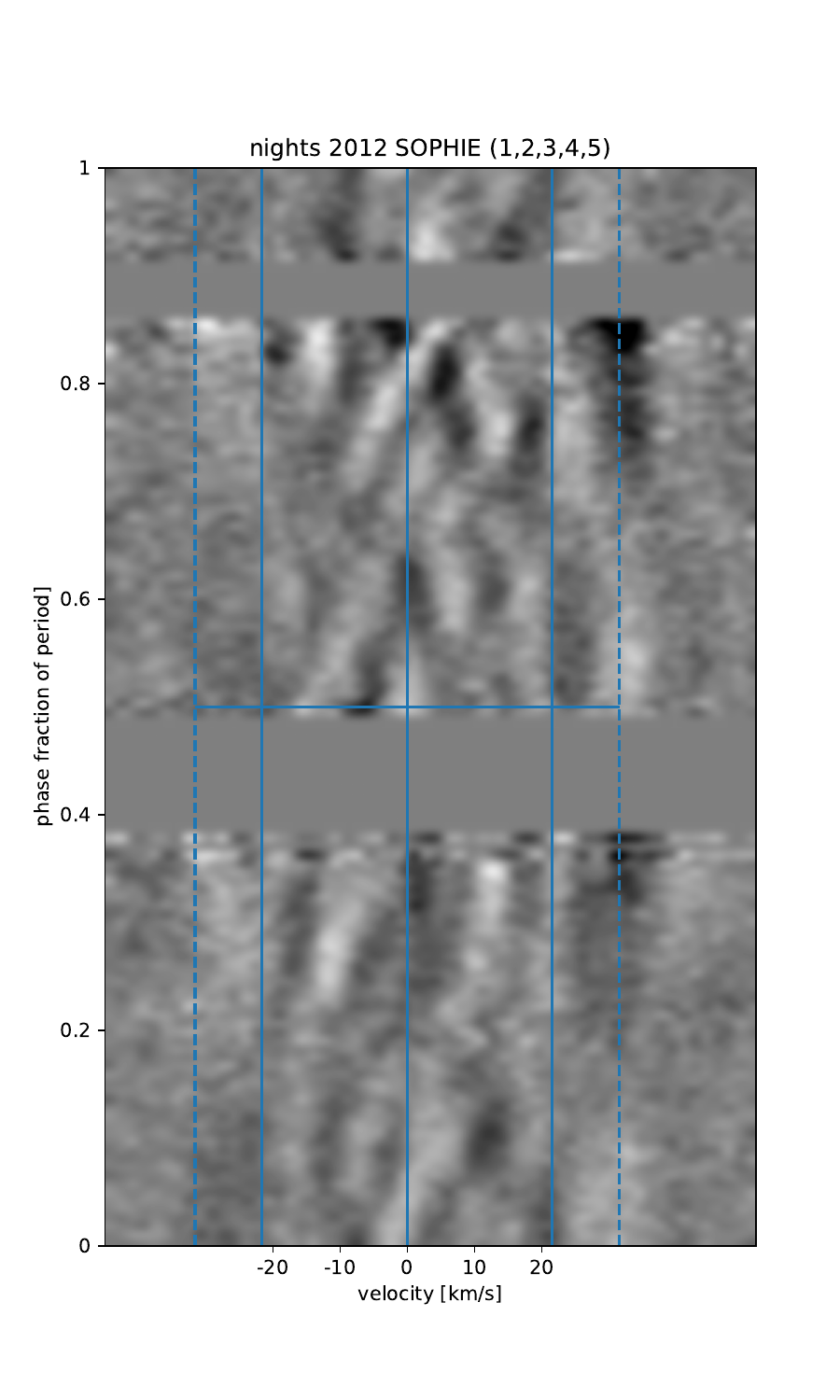}
\includegraphics[width=8cm]{vega_2025_figures/Sophie_2018_dynamic.pdf}
\includegraphics[width=9cm]{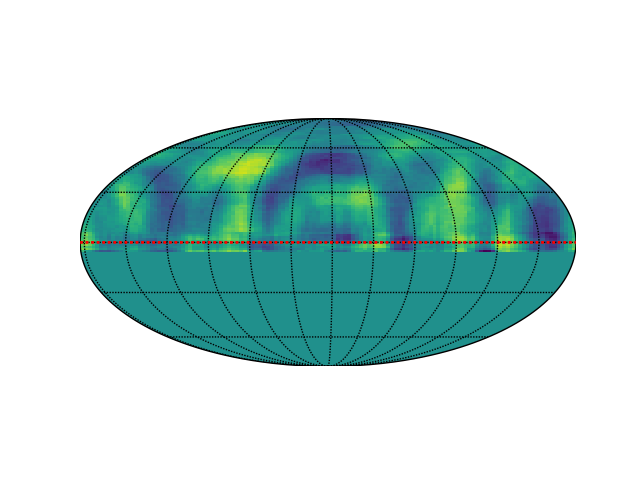}
\includegraphics[width=9cm]{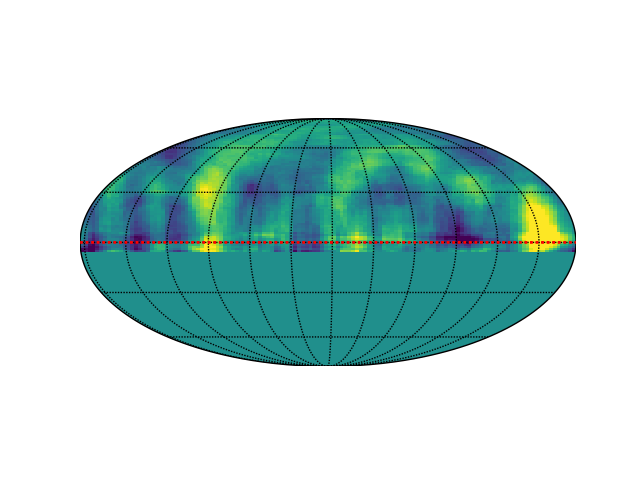}
\caption{\small Vega observations with SOPHIE/OHP in 2012 and 2018.  
Upper panels: dynamical spectra of 2012 and 2018. 
Phase zero corresponds for the 2018 data set to $BJD = 2456892.015$. while for the 2012 data set 
we opted for BJD = 2456142.332. The two inner blue lines symmetric to zero velocity correspond to $\pm\,\vsini= 21.6\,\kms$, 
while the outer dashed blue lines include the zone where gaussian broadening widens the line profile. Lower panels:
Reconstructed surface map of Vega in Mollweide projection, based on the procedure described in 
Sect.~\ref{dynamic}. Left map of 2012, right map of 2018. $80\%$ of the highest $S/N$ files were kept for map building.}
\label{vega_2012_2018}		
\end{figure*}

\begin{figure*} \centering
\includegraphics[width=8cm]{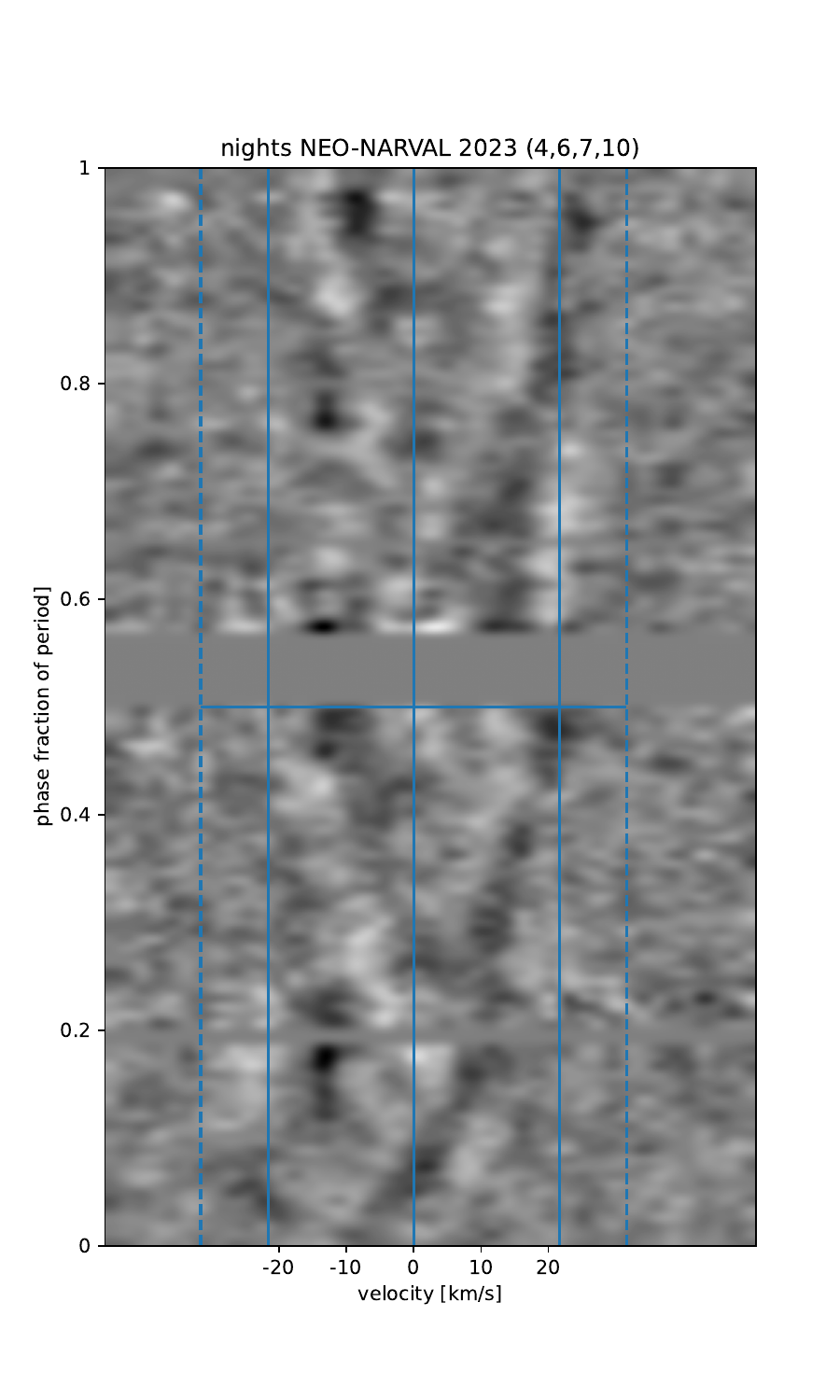}
\includegraphics[width=8cm]{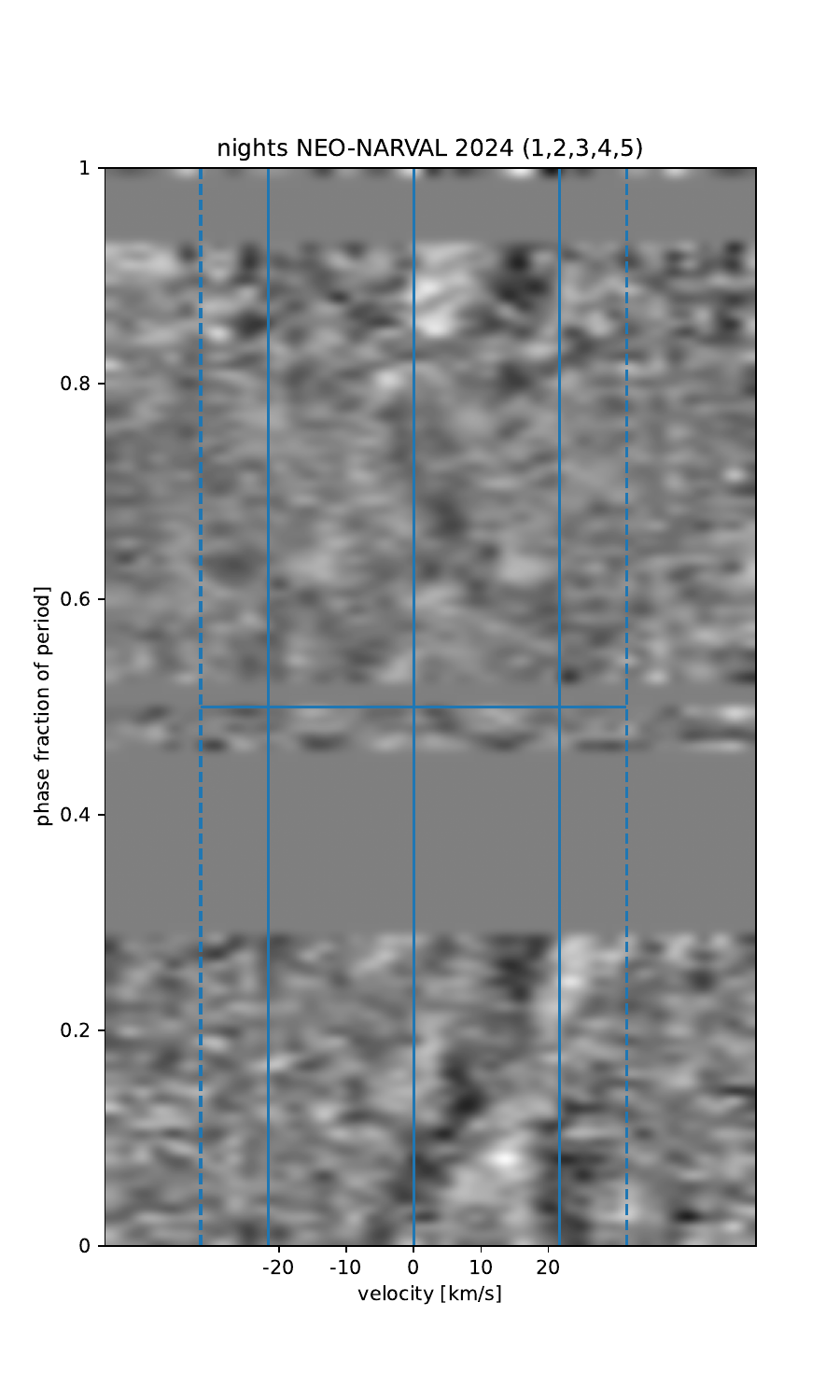}
\includegraphics[width=9cm]{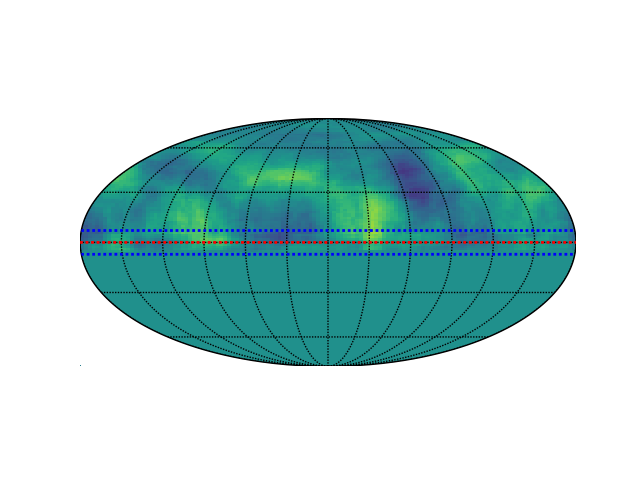}
\includegraphics[width=9cm]{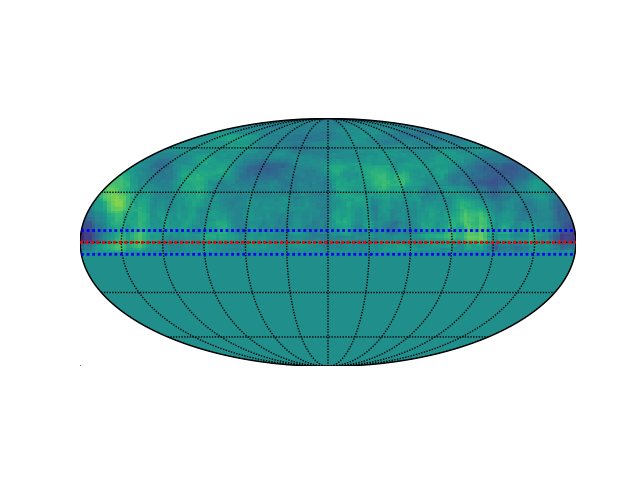}

\caption{\small Vega observations in 2023 (left) and 2024 (right) with NEO-NARVAL. See caption of~\ref{vega_2012_2018}	for details.}\label{vega_2023_2024}		
\end{figure*}

Before reconstructing the observed stellar maps we decided to exclude from the analysis the most noisy parts of the 2018 data set, i.e. between phases 0.265-0.328 and 0.742-0.828
of Fig.~\ref{vega_2012_2018} (upper, right). 
All reconstructed maps show the identical amplitude scale.
Figs~\ref{vega_2012_2018} and~\ref{vega_2023_2024} (lower panels) show the Mollweide projected reconstruction of 2012, 2018, 2023 and 2024 brightness maps. Bright stellar spots are represented in yellow on the reconstructed map. While in the 2012 data diffuse bright spots at different latitude seem to coexist, the 2018 map concentrates its bright zones above the equator, higher latitudes seem to be less covered by bright zones. This indicates a variation in the surface spot distribution over 6 years. The 2023 and 2024 maps seem to confirm the dominant location of activity spots nearby Vega's equator, despite the fact that the data sets were of lower quality. Still, the 2023 and 2024 maps are very different: the 2023 map shows two bright spots near the equator at phases 0.25 and 0.58,
while the 2024 data shows a brighter spot at approximate 30 degree latitude and close to phase 0. This tends to indicate spot variations within one year.

\begin{figure} \centering \includegraphics[width=9cm]{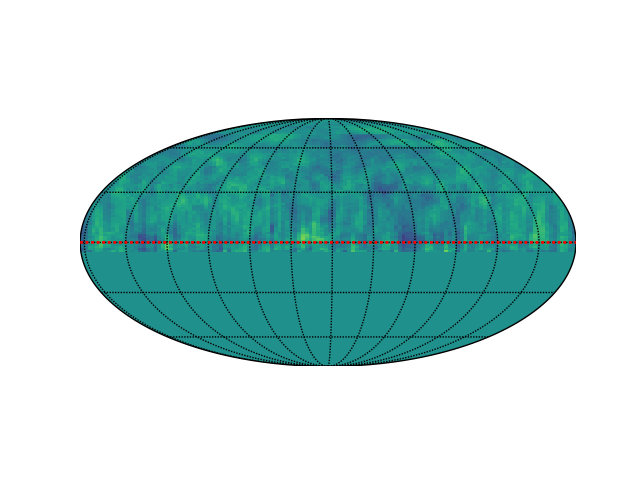}
\caption{\small Reconstructed surface map of a random noise dynamical spectrum. 
The standard deviation of the noise corresponds to the one measured outside the profile in the 2018 Sophie dynamical spectra.}
\label{vega_noise_spots}		 \
\end{figure}

In order to verify the reality of the reconstructed spots versus pure noise we
created an artificial dynamic spectrum with a random noise distribution ($\sigma
= 5 \times 10^{-5}$), similar to the pure noise outside the spectral profile in
the 2018 data set, the signatures in the dynamic spectra having overall
amplitudes of 3 $\times 10^{-4}$ F/F$_{c}$. Fig. \ref{vega_noise_spots}	clearly
shows that the reconstructed map only reveals fine fluctuations. A measurement
shows that the bright and dark signatures well above the 3\,$\sigma$ noise level
translate in the map representation in amplitude ranges at least twice above
those corresponding to pure noise.

\subsection{Search for a correlation between magnetic field and brightness maps}\label{search_corr}

A major motivation of the simultaneous observations with SOPHIE/OHP and NARVAL/TBL in 2018 was the analysis of magnetic versus activity spot location on the reconstructed stellar surface. First we used a simple visual inspection to  search for  similarities between the reconstructed starspot map Fig.~\ref{vega_2012_2018} (lower panel) and the surface magnetic field map Fig.~\ref{magneticmaps} from 2018.
 
Despite perfect alignment of the phase definition in the two representations, no obvious correspondence could be extracted (e.g. major activity spots located in equatorial regions with particularly strong radial or total magnetic field).

\begin{figure} \centering
\includegraphics[width=9cm]{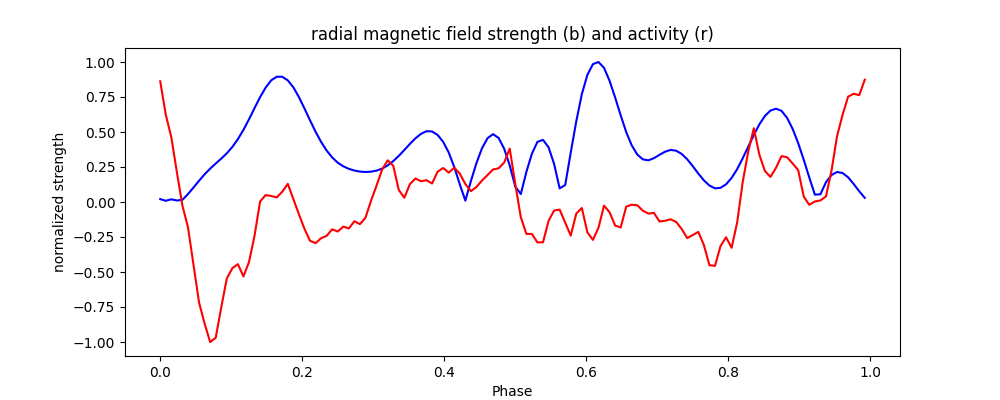}
\includegraphics[width=9cm]{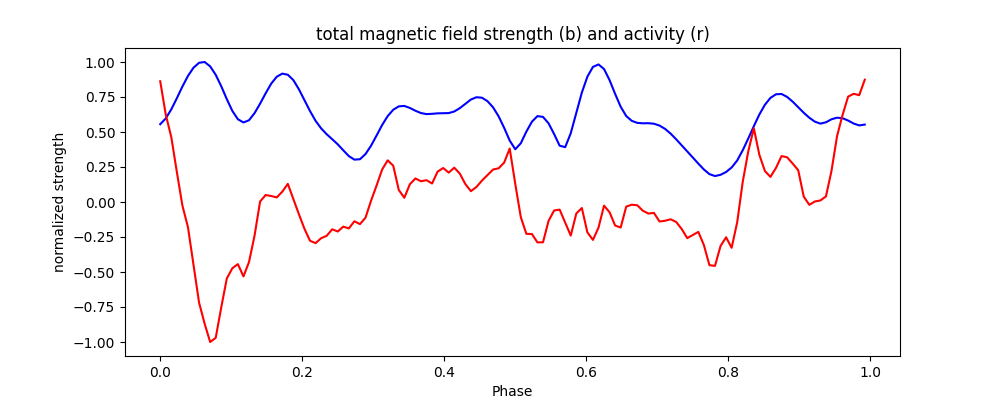}
\caption{\small Comparison of averaged latitudinal magnetic field strength variations (blue) and activity (red). Top: radial magnetic field and activity,
bottom: total magnetic field strength and activity.}
\label{mag_act}
\end{figure}

Since the dominant activity and magnetic features concentrate in regions near the equator, we decided to do a very rough one dimensional comparison of the location of magnetic versus activity spots in the enlarged equatorial belt of Vega, considering only latitudes between -7$^{\circ}$ and $45^{\circ}$, the lower edge of this range being the most extremes areas of the map reconstruction visible on the nearby pole-on star. To do so, we summed quantities in longitude (or phase) strips: i) the radial magnetic field strength, ii) the total magnetic field strength (square root of the quadradic addition of radial, meridional and azimuthal field strength) and iii) the starsport activity signature (intensity). This corresponds to project all information of the 2D maps on individual curves as a function of phase dependency only, and omitting latitude information. Three curves of averaged radial magnetic field strength, total magnetic field strength and activity signature as a function of phase were obtained and compared in Fig.~\ref{mag_act}.

At this stage no obvious correlation between these curves could be extracted, therefore no indication on correlated magnetic and activity spots has been demonstrated.

\section{Discussion and Conclusions} \label{disconc}

The aim of our work is to better understand the origin and evolution of Vega's faint observed stellar magnetic field. 
To do so, we are regularly monitoring Vega using high-resolution, high cadence spectroscopy and spectropolarimetry, 
in order to obtain complete sampling of several times Vega's short rotation period of $\approx$0.678\,d during one observing run.

In this work we managed successfully to cope with a major challenge, i.e. to connect new datasets (2023, 2024) 
obtained with a very different instrument (NEO-NARVAL) and a new data reduction pipeline (NEXTRA) to data obtained 
with NARVAL and reduced with the historic pipeline (LE), this despite the fact that magnetic signals are extremely 
faint in the case of Vega. We also demonstrated the robustness of our faint activity signatures by showing that they are identical in simultaneous data sets obtained with SOPHIE/OHP and NARVAL/TBL in 2018.

The spectropolarimetric observations of 2018, 2023 and 2024 obtained on Vega with NEO-NARVAL confirm the stability 
of the main magnetic signatures: a narrow polar spot of same average intensity during at least a decade and an 
inclined dipole indicating a localization nearby the equator and of comparable strength to previous observations. 
A stable magnetic field tends to support a fossil origin. However, small scale variations of the stellar magnetic field can not be excluded, due to S/N and resolution limitations.

While~\cite{boehm2015} reports the first detection of starspots on a hot star like Vega,~\cite{petit2017} 
re-analyzed the same SOPHIE data set from 2012 and obtained hints of surface features evolving over the different nights of the observing run. In the current work we were able to reveal long-term variations of the surface structure. 
A close comparison of our Mollweide projected maps of 2012 with those published in polar presentation by~\citep{petit2017} 
reveal strong similarities. 
Examining this line profile variability in three different ways leads to some consistent conclusions ($i$) The analysis of 
Fig.~\ref{ls_vspan_2018} shows the presence of a significantly larger number of harmonics in the Lomb-Scargle 
periodogram of $v_{\rm span}$. As shown in Fig.~6 in~\cite{boehm2015} this indicates a lower latitude of the spot location; ($ii$) 
a direct comparison of the dynamic spectra of 2012 and 2018 (shown in Fig.~\ref{vega_2012_2018}, upper panels) 
clearly shows that later data set has significantly stronger gradients of the spot signatures crossing the spectra, moving towards more extreme velocities; ($iii$) the reconstructed stellar surface maps (presented in Fig.~\ref{vega_2012_2018}, lower panels) shows higher latitude bright emission spots in the 2012 data set than in 2018. 
Obviously starspots concentrate in equatorial regions of Vega. 
It is also interesting to notice that the normalized amplitude of the signatures in all the dynamic spectra, from 2012 to 2024, are very similar.

Despite the very weak magnetic field signatures in Vega, the magnetic maps suggest that large scale surface magnetic field remains stable over more than a decade (probably even 16 years).  In contrast, the very weak intensity line profile variability most likely reveals surface spots that change on different time scales, an upper limit being the 6 years between the two SOPHIE data sets.

Under the assumption that starspots are linked to magnetism, variable starspots might be related 
to dynamo generated fields. We searched for similarities in spot distribution between the reconstructed starspot map of 2018 the surface magnetic field map, based on simultaneously obtained spectroscopic data at SOPHIE/OHP and spectropolarimetric data at NARVAL/TBL. While a final conclusion cannot be drawn from this limited dataset, it currently appears that the brightness spots are uncorrelated with the large-scale magnetic field.

The distribution and stability of the large scale magnetic field appears to be inconsistent with the distribution and variability of the brightness spots.  If the variability in the line intensity profiles is indeed due to brightness spots, and if these brightness spots are generated by magnetic fields as seen in cooler stars, then this discrepancy may reflect two different components of the stellar magnetic field. The first one would be a stable, large-scale component, contributing to the Stokes V signal. The second one is a variable, smaller scale component, which would be strong enough to produce brightness spots, but structured in a sufficiently complex pattern of mixed magnetic  polarities to mostly cancel out in Stokes V (a similar possibility was pointed out by \citealt{2020alhena} for the Am star Alhena). We can speculate that the two components of the magnetic field have different origins and highlight the coexistence of a weak fossil field embedded in the radiative portion of the star (providing the stable, large-scale structure), while a convective dynamo, most likely concentrated in the equatorial regions where the convective envelope is deeper, could produce the variable, smaller scale structure.  

Vega's "magnetic puzzle" is getting more mysterious with the insight of simultaneous brightness and magnetic surface reconstructions. Vega delivers, therefore, unique indications enabling us to obtain deeper insight into the magnetism of tepid/hot stars.

\begin{acknowledgements}
      The authors want to thank Alexis Lavail for fruitful discussions and support.
      CPF gratefully acknowledges funding from the European Union's Horizon Europe research and innovation 
      programme under grant agreement No. 101079231 (EXOHOST), and from the United Kingdom Research and Innovation (UKRI) 
      Horizon Europe Guarantee Scheme (grant number 10051045). The specpolFLow can be found at https://github.com/folsomcp/specpolFlow 
      and the LSDpy package at https://github.com/folsomcp/LSDpy. 
     Matthias Hoschneider acknowledges financing the support of the DFG CRC research grant 1248 at Potsdam university.
      
\end{acknowledgements}

\bibliographystyle{aa} 
\bibliography{vega_2025.bib} 

\end{document}